\documentclass[prd,twocolumn,twoside,preprintnumbers,superscriptaddress,nofootinbib,natbib,bm,amsrefs]{revtex4-2}
\usepackage{graphicx}% Include figure files
\usepackage{hyperref}% add hypertext capabilities
\usepackage{url}% add hypertext capabilities

\usepackage{xcolor}
\usepackage{amsmath, amstext, amsfonts,amssymb, amsthm}
\usepackage{comment}
\usepackage[utf8]{inputenc}
\usepackage{multirow}
\newcommand{\ov}{\overline}

\newcommand{\eg}{{\em e.g.\,}}

\newcommand{\GeV}{{\rm GeV}}

\newcommand{\TeV}{{\rm TeV}}

\newcommand{\cosmax}{{\cos\theta_{\rm max}}}
\newcommand{\cosmin}{{\cos\theta_{\rm min}}}

\newcommand{\sw}{\mathswitch {s_{\scrs\PW}}}

\newcommand{\PW}{\mathswitchr W}

\newcommand{\scrs}{\scriptscriptstyle}

\def\mathswitch#1{\relax\ifmmode#1\else$#1$\fi}
\def\mathswitchr#1{\relax\ifmmode{\mathrm{#1}}\else$\mathrm{#1}$\fi}

\definecolor{BlueViolet}{rgb}{0.2, 0.00, 0.7}
\definecolor{Blue}{rgb}{0.15, 0.00, 0.9}
\definecolor{lightblue}{rgb}{0.15, 0.35, 0.95}
\definecolor{kitgreen}{rgb}{0,
0.58823 %150/255
, 0.50980 %130/255
}

\newcommand{\Eprint}[1]{\href{#1}}

\definecolor{lb}{rgb}{.74,.83,.9}
\definecolor{ly}{rgb}{1,.92,.8}
\definecolor{lr}{rgb}{.98,.85,.87}

\begin{document}
\preprint{P3H-24-032, TTP24-014, KEK–TH–2622, DESY-24-072}
\title{Determining Weak Mixing Angle at $\mu$TRISTAN}

\author{Lisong Chen}
\email{lisong.chen@kit.edu}
\affiliation{Institute for Theoretical Particle Physics (TTP), Karlsruhe Institute of Technology (KIT), Wolfgang-Gaede-Str.\,1, 76131 Karlsruhe, Germany}
\author{Syuhei Iguro}
\email{igurosyuhei@gmail.com}
\affiliation{Institute for Theoretical Particle Physics (TTP), Karlsruhe Institute of Technology (KIT), Wolfgang-Gaede-Str.\,1, 76131 Karlsruhe, Germany}
\affiliation{Institute for Astroparticle Physics (IAP), KIT, Hermann-von-Helmholtz-Platz 1,\\ 76344 Eggenstein-Leopoldshafen, Germany}
\affiliation{Institute for Advanced Research (IAR), Nagoya University, Nagoya 464--8601, Japan}
\affiliation{Kobayashi-Maskawa Institute (KMI) for the Origin of Particles and the Universe,\\ Nagoya University, Nagoya 464--8602, Japan}
\affiliation{KEK Theory Center, IPNS, KEK, Tsukuba 305--0801, Japan}
\author{Yu Hamada}
\email{yuhamada@post.kek.jp}
\affiliation{Deutsches Elektronen-Synchrotron DESY, Notkestr. 85, 22607 Hamburg, Germany}
\affiliation{Research and Education Center for Natural Sciences, Keio University, 4-1-1 Hiyoshi, Yokohama, Kanagawa 223-8521, Japan}
\begin{abstract}
$\mu$TRISTAN is a realistic high energy lepton collider based on the existing technology aiming at indirect and direct search for the physics beyond the standard model (SM).
We propose a measurement to determine one of the most prominent parameters of the SM, weak mixing angle to test the SM and probe the new physics effect in M\o ller-like scattering with a wide range of interaction scales.
We show that this experiment not only can determine the weak mixing angle with percent to milli-level accuracy but also scan over a wide range of interaction scales that have never been archived in a single experiment elsewhere.
\end{abstract}
\maketitle

%
%
%

%%%%%%%%%%%%%%%%%%%%%%%%%%%%%%%%%%%%%
\section{Introduction}
\label{sec:intro}
%%%%%%%%%%%%%%%%%%%%%%%%%%%%%%%%%%%%%

Precision measurements of weak neutral current have been very important to verify the Standard Model of electroweak interactions; see Ref.\,\cite{PDG2022} for the recent review.
The ultra-precise measurements of the weak mixing angle at different interaction scales can test the electroweak theory and indirectly probe the new physics effect at the TeV scale.
Therefore, it provides a complementary probe to high-energy colliders, such as the Large Hadron Collider (LHC). 

The definition of a weak mixing angle depends on the renormalization scheme.
We discuss in the modified minimal subtraction scheme $\ov{\rm{MS}}$ in which quantum corrections appear as the scale dependence of the angle.
In this scheme, $\sin^2$ of the angle is given as 
\begin{align}
\sin^2\theta_w(\mu)=\frac{g^{\prime2}(\mu)}{g^2(\mu)+g^{\prime2}(\mu)},
\end{align}
where $g$ and $g^\prime$ respectively corresponds to the coupling constant of $SU(2)$ and $U(1)_Y$ of the SM.

At the very low energy $\simeq\mathcal{O}(10){\rm{MeV}}$, measuring nuclear weak charge with atomic parity-violating transitions can precisely determine $\sin^2 \theta_w$ ($s_W^2$ for short) \cite{Safronova:2017xyt}.
The heavy nuclei e.g. Cs, Ti, Pb, Ra, Yb, enjoy the large nuclear enhancement in parity violating effect, and further precision is foreseen \cite{Aubin:2013uwaa,Antypas:2017vni}. 
The parity-violating asymmetry $A_{PV}$ and polarization asymmetry $A_{LR}$ in t (and u)-channel topology e.g. $e^-e^- \to e^-e^-$ (M\o ller scattering) and $e^- p\to e^- p$ determine $s_W^2$ at $\simeq\mathcal{O}(10^{2\sim4}){\rm{MeV}}$ \cite{Prescott:1979dh,SLACE158:2005uay,Wang:2014guo,PVDIS:2014cmd,Qweak:2018tjf}.

On the other hand, a high-energy collider is also a powerful tool for the $s_W^2$ determination.
The NuTeV experiment determined $s_W^2$ with $\nu {\rm{N}}$ and $\ov\nu {\rm{N}}$ interactions using high energy $\nu$ and $\ov \nu$ beams \cite{NuTeV:2001whx}.
The previous precision measurements at the Z pole e.g. large electron positron (LEP) collider \cite{ALEPH:2005ab}, Stanford lepton collider (SLC) \cite{SLD:1996gjt,SLD:2000leq,SLD:2000jop,SLD:2000ujp}, and Tevatron \cite{D0:2017ekd,CDF:2018cnj} determined $s_W^2$ around $Q=m_Z$ based on forward-backward asymmetry $A_{ FB}$.
Recently the ATLAS, CMS, and LHCb also measured leptonic $A_{
FB}$ \cite{LHCb:2015jyu,ATLAS:2015ihy,ATLAS:2018gqq,CMS:2018ktx,CMS:2024aps} and the resulting precision is comparable with previous colliders. 
Also, the high luminosity (HL)-LHC which extends pseudorapidity acceptance will improve the statistics further and will halve the uncertainty  \cite{CMS:2017vxj,ATLAS:2018qvs,LHCb:2018roe} where remaining uncertainty largely comes from PDFs.
Moreover, there is a discrepancy between SLC and LEP, and the NuTeV result is not perfectly consistent with the SM prediction.\footnote{
Regarding the NuTeV result, the assumption of symmetric $s$-$\ov{s}$ parton distribution function, $\nu_e$ contamination from $K$ decays would account for the NuTeV discrepancy; however, the definitive answer remains elusive; see Ref.\,\cite{PDG2016} for detail.}
Therefore, further independent measurement is of crucial importance.

In this regard, a proposed $\mu^+ e^-$ and $\mu^+\mu^+$ collider, $\mu$TRISTAN could be interesting \cite{Hamada:2022mua}.\footnote{
There are several future measurements, DUNE \cite{deGouvea:2019wav}, Polarized Belle II \cite{USBelleIIGroup:2022qro}, FCC-ee \cite{FCC:2018evy}, CEPC \cite{Zhao:2022lyl}, ILC \cite{ILC:2013jhg}, Moller \cite{MOLLER:2014iki}, SoLID \cite{JeffersonLabSoLID:2022iod} and P2 \cite{Becker:2018ggl} which will measure $s_W^2$ at different scale.}
The clear advantage of this muon collider is that it can be readily realized based on the existing $\mu^+$ ultra cooling technology \cite{Hamada:2022mua}, which can not be applied with $\mu^-$. 
Not only can it reach the same and larger energy scales w.r.t. the future experiments listed in the footnote, but also have better sensitivity in testing Lepton Flavor Universality (LFU) since it uses $\mu^+$ beams in both proposed scatterings.
Although the primary targets of this collider are Higgs precision and direct search for heavy particles, 
this multi-purpose experiment is also useful for indirect searches for new physics since it is a lepton machine \cite{Hamada:2022uyn,Das:2022mmh,Fridell:2023gjx,Lichtenstein:2023iut,Dev:2023nha,Fukuda:2023yui,Okabe:2023esr,Goudelis:2023yni,Das:2024gfg,Heeck:2024uiz}.
For instance, the M\o ller-like scatterings of $\mu^+ \mu^+$ and $\mu^+ e^-$ exhibit nice sensitivity to deviations from the SM as studied within the SMEFT framework which would encode $\mathcal O(1\sim 10)$\, TeV new physics effect \cite{Hamada:2022uyn}.
We point out that, thanks to the well-controlled polarization of the $e^-$ beam and the large momentum of the $\mu^+$ beam, this $\mu^+ e^-$ M\o ller-like scattering is quite useful for the $s_W^2$ determination.
Moreover, utilizing the parity asymmetric component of the cross section, which is not taken in Ref.~\cite{Hamada:2022uyn}, one can reduce potential systematic uncertainty and determine $s_W^2$ precisely.
We evaluate the $\mu$TRISTAN sensitivity to $s_W^2$ in the following.

%% PRL version does not need an outline
The outline of this letter is given below.
We introduce the asymmetry in Sec.\,\ref{sec:ALR}.
In Sec.\,\ref{sec:pheno} we evaluate the $\mu$TRISTAN sensitivity of $s_W^2$. 
We conclude our findings in Sec.\,\ref{sec:conclusion} and briefly discuss the $\mu^+\mu^+$ version of $\mu$TRISTAN.

%%%%%%%%%%%%%%%%%%%%%%%%%%%%%%%%%%%%%
\section{Polarization asymmetry}
\label{sec:ALR}
%%%%%%%%%%%%%%%%%%%%%%%%%%%%%%%%%%%%%

At the proposed $\mu^+ e^-$ collider, t-channel gauge boson exchange allows us to access $s_W^2$.
We consider the collision in the center of mass frame
$p_{\mu,i}=(E,0,0,E)$, $p_{e,i}=(E,0,0,-E)$, $p_{\mu,f}=(E,0,E\sin\theta,E\cos\theta)$ and $p_{e,f}=(E,0,-E\sin\theta,-E\cos\theta)$
with 
$s=4E^2=4E_\mu E_e$, $t=-2 E^2(1-\cos\theta)$, $u=-s-t$.
$\theta$ is the opening angle in the center of mass (CM) frame, and
$E_{\mu}$ and $E_{e}$ are initial beam energy in the lab frame.\footnote{See Appendix \ref{App:kin} for the relation between the CM frame and lab frame.}
We neglected the charged lepton masses.

We consider the polarization asymmetries defined in terms of differential cross sections as  
\begin{widetext}
\begin{align}
A_{PV}^e(P_{e^-},\cos\theta_{\rm min},\cos\theta_{\rm max})
&=\int_{\cos\theta_{\rm min}}^{\cos\theta_{\rm max}} d\cos\theta \left(\frac{d\sigma_{0,P_{e^-}} }{d\cos\theta} -\frac{d\sigma_{0,-P_{e^-} } }{d\cos\theta} \right) 
\biggl/\left(\frac{d\sigma_{0,P_{e^-}} }{d\cos\theta}+\frac{d\sigma_{0,-P_{e^-}}}{d\cos\theta} \right),\\
A_{PV}^\mu
(P_{\mu^+},\cos\theta_{\rm min},\cos\theta_{\rm max})
&=\int_{\cos\theta_{\rm min}}^{\cos\theta_{\rm max}} d\cos\theta\left(\frac{d\sigma_{P_{\mu^+},0} }{d\cos\theta} -\frac{d\sigma_{0,0 } }{d\cos\theta} \right) 
\biggl/\left(\frac{d\sigma_{P_{\mu^+},0} }{d\cos\theta}+\frac{d\sigma_{0,0}}{d\cos\theta} \right).
\end{align}
The polarized cross section is subdivided as

\begin{align}
\frac{d\sigma_{P_{\mu^+},P_{e^-}}}{d\cos\theta}=
\frac{1}{4}\biggl{\{}& \left(1-P_{\mu^+}\right) \left(1+P_{e^-}\right)\frac{d \sigma_{LR}}{d \cos\theta}
+\left(1+P_{\mu^+}\right) \left(1-P_{e^-}\right)\frac{d \sigma_{RL}}{d\cos \theta} \nonumber\\
 &+\left(1+P_{\mu^+}\right) \left(1+P_{e^-}\right)\frac{d \sigma_{RR}}{d \cos\theta}
 +\left(1-P_{\mu^+}\right) \left(1-P_{e^-}\right)\frac{d \sigma_{LL}}{d \cos\theta}\biggl{\}},
\end{align}

%\si{
%\begin{align}
%\frac{d\sigma_{P_{\mu^+},P_{e^-}}}{d\cos\theta}=
%\frac{1}{4}\biggl{\{}& \left(1-P_{\mu^+}\right) \left(1+P_{e^-}\right)\frac{d \sigma_{RR}}{d \cos\theta}
%+\left(1+P_{\mu^+}\right) \left(1-P_{e^-}\right)\frac{d \sigma_{LL}}{d\cos \theta} \nonumber\\
% &+\left(1+P_{\mu^+}\right) \left(1+P_{e^-}\right)\frac{d \sigma_{LR}}{d \cos\theta}
% +\left(1-P_{\mu^+}\right) \left(1-P_{e^-}\right)\frac{d \sigma_{RL}}{d \cos\theta}\biggl{\}},
%\end{align}
%}

where the polarization of a right (left) handed fermion corresponds to $P_{f}=+1\,(-1)$ and the polarization of a right (left) handed anti-fermion corresponds to $P_{\ov{f}}=-1\,(+1)$. 
The subscript ``$0$'' means unpolarized beam, $P_{e^-/\mu^+}=0$.
Assuming the SM, each differential polarized cross section at the tree level is given as

\begin{align}
    \frac{d \sigma_{LL}}{d \cos\theta}=\frac{d \sigma_{RR}}{d \cos\theta}=\frac{ s }{8\pi}\left|\sum_{V=\gamma,Z}\frac{g_L^{l,V} g_R^{l,V}}{t-m_V^2}\right|^2,~~
    \frac{d \sigma_{LR}}{d \cos\theta}=\frac{u^2}{8\pi s}\left|\sum_{V=\gamma,Z}\frac{g_R^{l,V} g_R^{l,V}}{t-m_V^2}\right|^2,~~
    \frac{d \sigma_{RL}}{d \cos\theta}=\frac{u^2}{8\pi s}\left|\sum_{V=\gamma,Z}\frac{g_L^{l,V} g_L^{l,V}}{t-m_V^2}\right|^2,
\end{align}

%\si{
%\begin{align}
%    \frac{d \sigma_{LR}}{d \cos\theta}=\frac{d \sigma_{RL}}{d \cos\theta}=\frac{ s }%{8\pi}\left|\sum_{V=\gamma,Z}\frac{g_L^{l,V} g_R^{l,V}}{t-m_V^2}\right|^2,~~
%
%    \frac{d \sigma_{RR}}{d \cos\theta}=\frac{u^2}{8\pi s}\left|\sum_{V=\gamma,Z}\frac{g_R^{l,V} g_R^{l,V}}{t-m_V^2}\right|^2,~~
%
%    \frac{d \sigma_{LL}}{d \cos\theta}=\frac{u^2}{8\pi s}\left|\sum_{V=\gamma,Z}\frac{g_L^{l,V} g_L^{l,V}}{t-m_V^2}\right|^2,
%\end{align}
%}
\end{widetext}
where $g_L^{l,\gamma}=g_R^{l,\gamma}=eQ_f$, $g_L^{l,Z}=\frac{g}{c_W}\left(T^3_l-Q_l s_W^2\right)$ and $g_R^{l,Z}=-\frac{g}{c_W} Q_l s_W^2$ with $Q_l=-1$, $T^3_l=-1/2$ are defined.
Indices $i,\,j$ of $\sigma_{ij}$ denote the beam polarizations of initial $\mu^+$ and $e^-$ beams, respectively.
The lepton flavor universal interaction of the SM is assumed.
The energy transfer in the t-channel propagator, $Q^2\equiv-t$ is a function of $\cos \theta$ for given beam energy, and hence scanning $\cos \theta$ corresponds to scanning $\mu$.
With s-channel topology \eg LEP and LHC, the reconstructing Z-boson mass uniquely determined $Q$ and other t-channel experiments \eg SLAC E158 experiment worked with the far finite opening angle detector which fixes $Q$.

%%%%%%%%%%%%%%%%%%%%%%%%%%%%%%%%%%%%%%%%%%%%%%%%%%
\begin{figure*}[t]
\centering
    \includegraphics[width=0.46\textwidth]{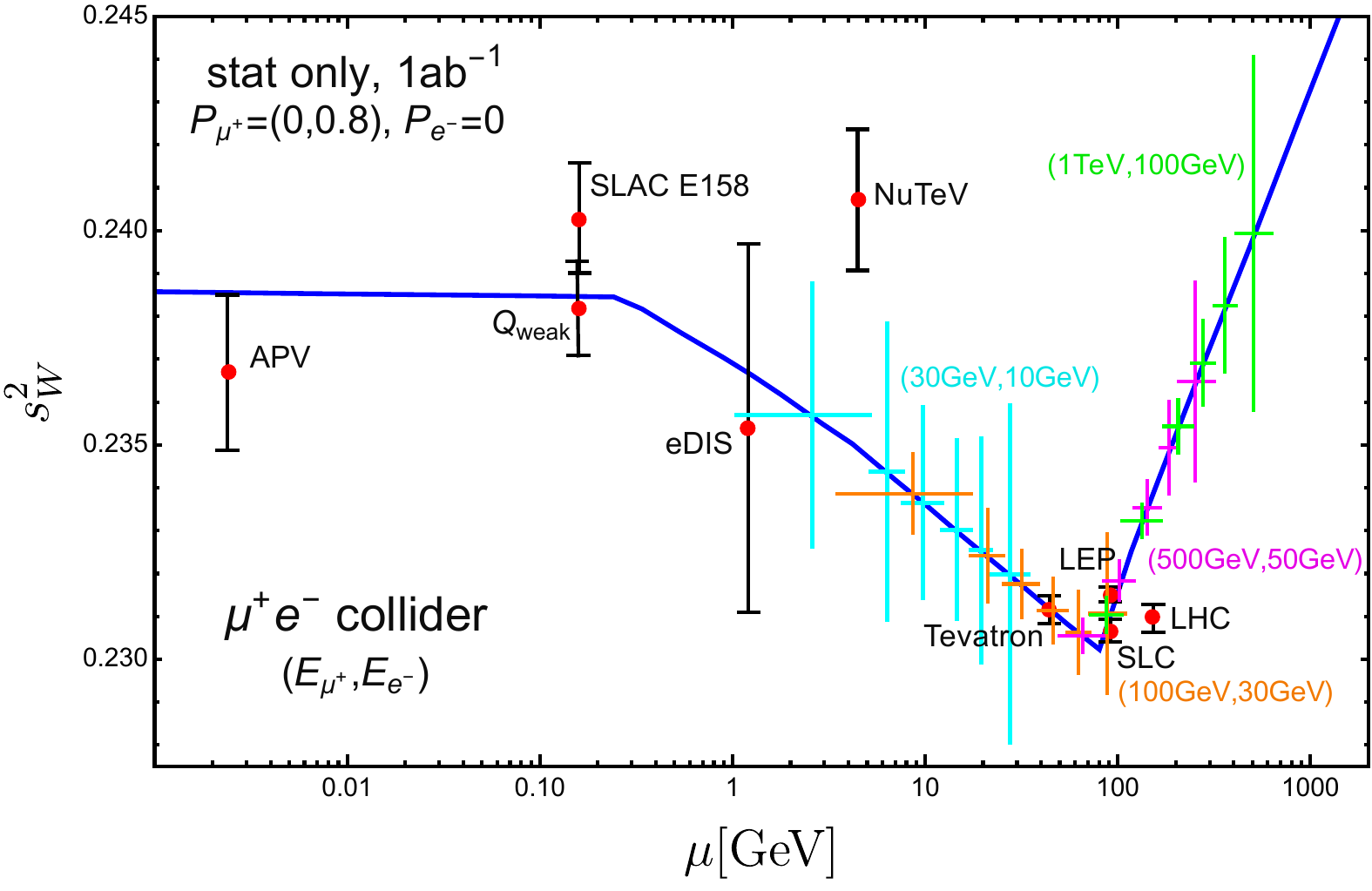}~~~~~~
    \includegraphics[width=0.46\textwidth]{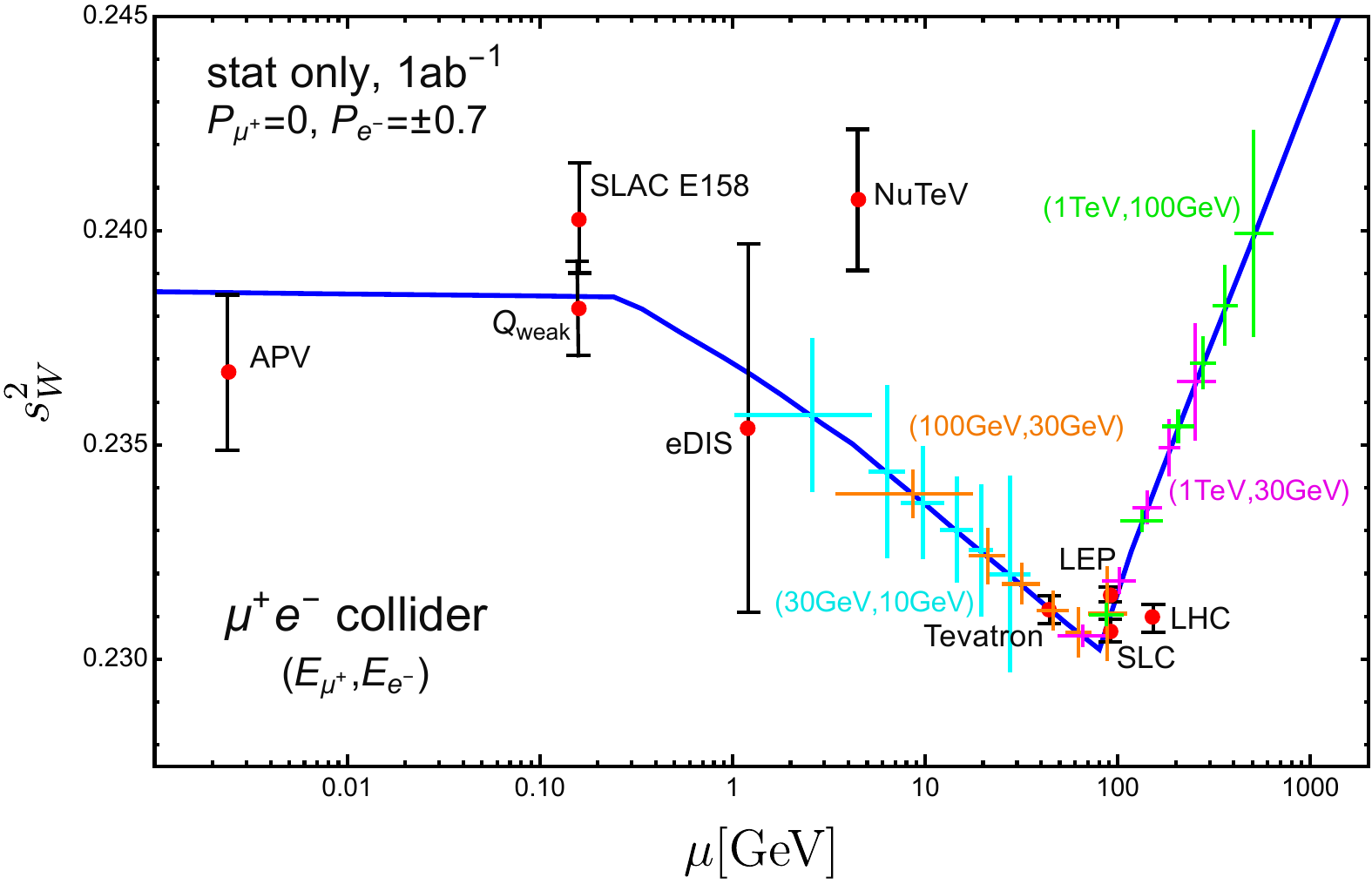}
    \caption{ 
    $\mu$TRISTAN sensitivity to $s_W^2$ with the polarized muon beam (left) and electron beam (right). 
    The statistical uncertainty is shown in the size of the vertical bar.
    The blue contour shows the SM prediction.
    Colored bins correspond to the $\mu$TRISTAN sensitivity for a given set of initial beam energy assuming 1ab$^{-1}$ of data with the polarized $\mu^+$.
    The existing constraints are shown with 1$\sigma$ error.
    Results from Tevatron, LEP, SLC, and LHC are all around $\mu=m_Z$, however, they are slightly shifted for better readability.
    Strictly speaking, the reference energy scale $Q_{ref}$ in our measurement is different from the scale $\mu$ on which $s_W$ depends in the $\overline{MS}$ scheme, which means the sensitivity study at the $\mu$TRISTAN features no running effect.
    Nevertheless, we identify them by overlaying the colored bins on the blue-band for the sake of illustration. 
    }
    \label{fig:emu}
\end{figure*}
%%%%%%%%%%%%%%%%%%%%%%%%%%%%%%%%%%%%%%%%%%%%%%%%%%

It is noted that the higher-order up to NNLO EW radiative corrections shift the parity-violating asymmetries by roughly $41.3\%$ mainly due to the $\gamma-Z$ mixing via hadronic vacuum polarization \cite{Czarnecki:1995fw, Du:2019evk,Erler:2022ckm}, resulting in approximately $3\%$ shift on $s_W^2$.
The MC tools such as {\sc SANC}  \cite{Andonov:2004hi,Arbuzov:2021oxs} will also help us have systematic uncertainty under better control. 
A thorough analysis would require the same amount of running effect given by radiative corrections. Certainly, the precision study of $\sw$ at $\mu$TRISTAN, although out of the scope in this paper, is indispensable, given the projected sensitivity, which will be discussed in the following section. 
We leave the thorough inspection of the precision study to future projects.

%%%%%%%%%%%%%%%%%%%%%%%%%%
\section{$\mu$TRISTAN sensitivity}
\label{sec:pheno}
%%%%%%%%%%%%%%%%%%%%%%%%%%

At this stage, there is no concrete detector design for $\mu$TRISTAN. The HL-LHC detector is assumed \cite{Kaji} as a benchmark study, although $\mu$TRISTAN is a lepton collider. 
We expect a better control of uncertainty at a future lepton collider.\footnote{We note that the energy resolution of the future lepton collider, e.g., ILC detector, is a factor 2 better for $\eta<2.5$ compared to the HL-LHC detector, but unfortunately, it is not available for $\eta\ge3.0$ currently.
In reality, the opening angle of the detector should be asymmetric, e.g., $-4 < \eta <2 $.
This is because we need a shield to protect the detector from decay products of the beam muons, which depends on muon beam energy.
When we assume the active opening angle of $15^\circ < \theta^\prime < 165^\circ$ ($10^\circ < \theta^\prime < 170^\circ$) to capture both final state particles where $\theta^\prime$ is the opening angle in the lab frame, we will lose the two lowest $\mu$ (one) bins in Fig.\,\ref{fig:emu}
while the impact on the large $\mu$ bins is negligible. 
A more detailed experimental study is necessary to address these issues.}
The HL-LHC detector covers up to $|\eta|=4$, approximately corresponding to $\theta^\prime > 2^\circ$ where $\theta^\prime$ is the opening angle in the lab frame.
As shown in Appendix\,\ref{sec:app_stat}, the statistical uncertainty of $s_W^2$ is expressed by 
\begin{align}
    \delta s_W^2 = \left(\frac{\delta A_{\rm LR}} {\delta s_W^2} \right )^{-1}\frac{1}{\sqrt{N_{\rm{tot}}}}.
\end{align}
We assume that the systematic error is small compared to the statistical one.
Unlike the LHC case, the asymmetry is reconstructed by the detector at the common angle, which mostly cancels out uncertainty.
The precise estimation of the systematic errors needs the careful consideration of experimental setup \cite{OPAL:2004xxz,Delahaye:2019omf,ZurbanoFernandez:2020cco,ILC2022}, and hence we simply neglect this at this stage.\footnote{Nevertheless we will discuss the potential uncertainty in Appendix \ref{sec:app_stat}.} 
We divide the opening angle into five to six sub-segments.
The width of $\cos\theta$ is set such that the detector resolution is smaller than each width.
The electron and muon tagging efficiencies are assumed to be $100\%$.

Based on the event number for the given $\cos\theta$ range and assuming the SM we can calculate the statistic uncertainty of $s_W^2$.
We defined a reference (event rate weighted) momentum scale $\langle Q_{ref} \rangle \equiv \int^{\cosmax}_{\cosmin} d\cos\theta\,Q\times 
\sqrt{ \sum_{ij} \frac{d\sigma_{ij}} {d\cos\theta} }\biggl/\int^{\cosmax}_{\cosmin} d\cos\theta \sqrt{\sum_{ij} \frac{d\sigma_{ij} }{d\cos\theta}}$ to recast the sensitivity on the $s_W^2$ vs $\mu$ plane.
Regarding polarization we fixed $P_{e^-}=\pm0.7$ and $P_{\mu^+}=0,\,+0.8$ \cite{Roney:2021pwz,Hamada:2022mua}.

At this stage of the $\mu$TRISTAN project, it would be interesting to consider several energy beam combinations to see the potential.
In Fig.\,\ref{fig:emu} (left) we considered four cases as a benchmark setup: $(E_\mu,\,E_e)=(1\TeV,\,100\GeV)$,\,$(500\GeV,\,50\GeV)$,\,$(100\GeV,\,30\GeV)$, and $(30\GeV,\,10\GeV)$ in green, magenta, orange, and cyan for each with $P_{\mu^+}=0,+0.8$ and $P_{e^-}=0$. 
The integrated luminosity of 1\,ab$^{-1}$ is assumed for each.
Fig.\,\ref{fig:emu} (right) shows the projected sensitivity with $P_{\mu^+}=0$ and $P_{e^-}=\pm0.7$.

For the wide range of $\mu$, these future colliders can precisely determine $s_W^2$ with considerable accuracy.
The larger $\mu$ bins suffer from the small event number which inflates the uncertainty.
Moreover, the lowest $\mu$ bin tends to have the larger uncertainty due to the small $A_{PV}$.
Interestingly, the high energy options (magenta and green) allow us to scan $s_W^2$ at $\mu\ge m_Z$.
On the other hand, the low energy ones (orange and cyan) can validate the intermediate region of $\mu\simeq10\sim 100\,$GeV. 
Therefore we conclude that those future lepton colliders provide a crucial precision test of the SM and $\mathcal O(1\sim 10)$ TeV scale new physics.

%%%%%%%%%%%%%%%%%%%%%%%%%%%%%%%%%%%%%
\section{Conclusions and discussion}
\label{sec:conclusion}
%%%%%%%%%%%%%%%%%%%%%%%%%%%%%%%%%%%%%
In this letter, we evaluated the $\mu$TRISTAN sensitivity to the prominent SM parameter, weak mixing angle at different scales.
The $\mu$TRISTAN is one of the most promising future Higgs factories that can be readily built based on the existing $\mu^+$ ultra cooling technology.
The precisely determined $s_W^2$ at Z pole, even though a more than $3\sigma$ discrepancy between SLAC and LEP has been reported, can be used to indirectly probe the new physics at TeV scale which affects the $s_W^2$ via RG running.
One obvious advantage given by $\mu$TRISTAN is that it gives a better constraint on non-LFU models due to the nature of $\mu^+\,e^-$ scattering.
Thanks to the nature of the t-channel process and the large $\sqrt{s}$ of $\mathcal {O}(10^{2}\sim10^{4})$\,GeV, we can scan $s_W^2$ over the wide range of interaction scale by reconstructing the scattering angle.
We found that $\mu$TRISTAN can not only measure $s_W^2$ at $\mu\ge m_Z$ for the first time but also allows us to scan $s_W^2$ down to $\mathcal{O}(1)\,\GeV$ with considerable accuracy.

It should be repeated that 
since the study here is based on the simplified analysis e.g., leading order process without initial/final state radiation and without a concrete detector design, 
a more detailed study is necessary from both the theoretical and experimental sides.

Although we focused on the $\mu^+\,e^-$ option of $\mu$TRISTAN in this letter, the other option $\mu^+\,\mu^+$ collider is also interesting to measure $s_W^2$ at a different scale.
In this setup, the final state is identical and hence indistinguishable.
As a result, both t-channel and u-channel prevent us from uniquely determining $Q$.
Nevertheless, it would be interesting to point out that at $\cos\theta=0$ the transfer energies in both channels coincide.
Therefore it would be possible to measure $s_W^2$ by changing $\sqrt{s}$. \\

%%%%%%%%%%%%%%%%%%%%%%%%%%%%%%%%%%%%%
\begin{acknowledgments} 
We thank Yasuyuki Horii, Ryuichiro Kitano, Surabhi Tiwari, and Xunwu Zuo for enlightening discussions. We thank Ayres Freitas for reading the manuscript and giving valuable feedback.
We appreciate Robert Ziegler for successfully organizing the BSM seminar at KIT, which motivated this work.
S.\,I. and L.\,C. are supported by the Deutsche Forschungsgemeinschaft (DFG, German Research Foundation) under grant 396021762-TRR\,257.
S.\,I. is also supported by the JSPS  Core-to-Core Program (Grant No.\,JPJSCCA20200002).
Y.\,H. is supported by JSPS Grant-in-Aid for Scientific Research KAKENHI Grant No.~JP22KJ3123 
and the Deutsche Forschungsgemeinschaft under Germany's Excellence Strategy - EXC 2121 Quantum Universe - 390833306.
\end{acknowledgments}
%%%%%%%%%%%%%%%%%%%%%%%%%%%%%%%%%%%%%

\begin{widetext}
\appendix

%%%%%%%%%%%%%%%%%%%%%%%%%%%%%%
\section{Kinematics}
\label{App:kin}
%%%%%%%%%%%%%%%%%%%%%%%%%%%%%
In the main text, we consider the CM frame.
Here we will show the relationship between the CM frame and the lab frame.
In the collider we consider energetic $\mu^+$ injection with $p_1=(E_{\mu,i},\,0,\,0,\,E_{\mu,i})$ which collides with electron beam of $p_2=(E_{e,i},\,0,\,0,\,-E_{e,i})$ resulting in $p_3=(E_{\mu,f},0,E_{\mu,f}\sin\theta^\prime,E_{\mu,f}\cos\theta^\prime)$ and $p_4=(E_{\mu,i}+E_{e,i}-E_{\mu,f},0,-E_{\mu,f}\sin\theta^\prime, E_{\mu,i}-E_{e,i}-E_{\mu,f}\cos\theta^\prime)$ where masses are neglected and $E_{\mu,f}= 2E_{\mu,i} E_{e,i}/(E_{\mu,i}+E_{e,i}-(E_{\mu,i}-E_{e,i})\cos\theta^\prime)$.
$\theta^\prime$ is the opening angle between initial and final $\mu$ in the lab frame.
However, a calculation of cross section is easy in the CM frame, which we defined at the beginning of Sec.\,\ref{sec:pheno}.
The connecting relation of angles is given as
\begin{align}
    \cos\theta^\prime=\frac{(E_{\mu,i}+E_{e,i})\cos\theta +E_{\mu,i}-E_{e,i}}{E_{\mu,i}+E_{e,i} +(E_{\mu,i}-E_{e,i})\cos\theta}.
\end{align}

%%%%%%%%%%%%%%%%%%%%%%%%%%%%%%%%%%%%%
\section{Uncertainty}
\label{sec:app_stat}
%%%%%%%%%%%%%%%%%%%%%%%%%%%%%%%%%%%%%
The uncertainty in a generic function $F(x_1,\,x_2,\,x_3,....x_n)$ with $x_i$ being the argument of $F$ is given as
\begin{align}
    \delta F(x_1,\,x_2,\,x_3,....x_n)=\sqrt{\sum_i^n \left( \frac{d F}{d x_i}\right)^2(\delta x_i)^2}.
\end{align}
We evaluate the statistic uncertainty of asymmetry $\mathcal{A}$ which is the function of polarized cross sections, $\sigma_L$ and $\sigma_R$ as,
\begin{align}
    \mathcal{A}=\frac{\sigma_R-\sigma_L}{\sigma_R+\sigma_L}=\frac{N_R-N_L}{N_R+N_L}
\end{align}
where $\sigma$ and $N$, respectively, denote the cross section and the number of events for a given polarized beam for a given opening angle $\theta^\prime$.
The derivatives respective to $N_R$ and $N_L$ are respectively given as
\begin{align}
  \frac{d \mathcal{A}} {d N_R} = \frac{2 N_L}{(N_L+N_R)^2},\,\,  \frac{d \mathcal{A}} {d N_R} = -\frac{2 N_R}{(N_L+N_R)^2},
\end{align}
and hence we obtain 
\begin{align}
    \delta \mathcal{A} = \frac{2}{(N_R+N_L)^2} \sqrt{N_L^2 (\delta N_R)^2+N_R^2 (\delta N_L)^2}.
\end{align}
When $\mathcal{A} \ll 1$ holds we can simplify the expression as 
\begin{align}
    \delta \mathcal{A} \simeq \frac{\delta N_R}{\sqrt{2} N_R}=\frac{\delta N_R^r}{\sqrt{2}}=\frac{1}{\sqrt{2}}\frac{1}{\sqrt{N_R}}=\frac{1}{\sqrt{N_{\rm{tot}}}},
\end{align}
where $N_R\simeq N_L\simeq N_{\rm{tot}}/2$ is used.
Based on the chain rule and assuming the other uncertainty is negligible, we evaluate the uncertainty of $s_W^2$ as  
\begin{align}
    \delta s_W^2 = \left(\frac{\delta \mathcal{A}} {\delta s_W^2} \right )^{-1} \delta \mathcal{A}.
\end{align}
We evaluate the uncertainty at $\langle Q_{ref} \rangle$ to overlay on the plane.

The uncertainty in initial beam energy has only a negligible impact. 
Similarly, the uncertainty in the electron beam polarization is known to be $0.25\sim0.5\%$ or better \cite{Roney:2021pwz,Vormwald:2015hla,ILCInternationalDevelopmentTeam:2022izu} and has a negligible impact on $s_W^2$ determination.
On the other hand, we have a larger uncertainty in the polarization of the polarized muon beam.
For instance, the $5\%$ shift of polarization results in the $5\%$ shift in the asymmetry.
This corresponds to a shift of $2.5\sim3.5$ $\times 10^{-3}$ in $s_W^2$, depending on the beam energy configuration.
Therefore to tame the uncertainty such that the associated uncertainty is smaller than the statistical one in Fig.\,\ref{fig:emu} (left), we need a few $\%$ determinations of the muon polarization.
In $\mu$TRISTAN, each bunch has $\mathcal{O}(10^{10})$ $\mu^+$ and its life time $\tau_\mu$ is $\mathcal{O}(10)\,$ms.
We expect $10^3$ decaying muon per meter per bunch and hence a polarimeter of several meters can achieve $\mathcal{O}(1)\%$ determination \cite{Norum:1996mi}.

\end{widetext}
%%%%%%%%%%%%%%%%%%%%%%%%%%%%%%%%%%%%%%%%%%%
\bibliographystyle{utphys}
\bibliography{references}
\end{document}